%% file: main.tex
\documentclass[sigconf]{acmart}
\usepackage[utf8]{inputenc}
\usepackage{graphicx}
\usepackage{subcaption}
\usepackage{xcolor}
\usepackage{xspace}
\usepackage{todonotes}
\usepackage{hyperref}
\usepackage{framed, mdframed}
\usepackage{multicol}
\usepackage{balance}
\usepackage{enumitem}
\usepackage{float}
\usepackage{appendix}
\usepackage{wrapfig,booktabs}
\usepackage{hyperref}
\usepackage{url}

\renewenvironment{quote}{%
   \list{}{%
     \leftmargin0.45cm   
     \rightmargin\leftmargin
   }
   \item\relax
}
{\endlist}

\usepackage{xspace,xpunctuate}
\newcommand{\ie}{{i.e.,}\xspace}
\newcommand{\eg}{{e.g.,}\xspace}

\newcommand{\etc}{{etc\xperiod}\xspace}

\newcommand{\system}{network report\xspace}
\newcommand{\System}{Network report\xspace}

\newcommand{\systems}{network reports\xspace}

\AtBeginDocument{%
  \providecommand\BibTeX{{%
    \normalfont B\kern-0.5em{\scshape i\kern-0.25em b}\kern-0.8em\TeX}}}

\setcopyright{acmcopyright}
\copyrightyear{2022}
\acmYear{2022}
\acmDOI{XXXXXXX.XXXXXXX}

\begin{document}
\setlength{\belowdisplayskip}{2pt}
\setlength{\abovedisplayskip}{2pt}
\title{Network Report: A Structured Description for Network Datasets}

\author{Xinyi Zheng}
\affiliation{%
  \institution{Carnegie Mellon University}\country{}
}

\author{Ryan A. Rossi}
\affiliation{
\institution{Adobe Research}\country{}}

\author{Nesreen Ahmed}
\affiliation{
\institution{Intel Labs}\country{}}

\author{Dominik Moritz}
\affiliation{
\institution{Carnegie Mellon University}\country{}}


\begin{abstract}
The rapid development of network science and technologies depends on shareable datasets.
Currently, there is no standard practice for reporting and sharing network datasets.
Some network dataset providers only share links, while others provide some contexts or basic statistics.
As a result, critical information may be unintentionally dropped, and network dataset consumers may misunderstand or overlook critical aspects. 
Inappropriately using a network dataset can lead to severe consequences (\eg discrimination) especially when machine learning models on networks are deployed in high-stake domains. Challenges arise as networks are often used across different domains (\eg network science, physics, etc) and have complex structures. 
To facilitate the communication between network dataset providers and consumers, we propose \system.
A \system is a structured description that summarizes and contextualizes a network dataset. 
\System extends the idea of dataset reports (\eg Datasheets for Datasets) from prior work with network-specific descriptions of the non-i.i.d. nature, demographic information, network characteristics, \etc
We hope \systems encourage transparency and accountability in network research and development across different fields. 
\end{abstract}

\maketitle

\section{Introduction}
Scientists use networks as representation to understand and model complex systems including social, biological, neural, and technological systems~\cite{Newman2003TheSA}.
Networks are also at the heart of industrial applications, empowering products from Google~\cite{Page1999ThePC} to Facebook, CISCO, and Twitter.
Recently, machine (deep) learning on graphs and networks has received increasing attention in major conferences, driving real-world applications such as drug discovery~\cite{You2018GraphCP}, recommender systems~\cite{Ying2018GraphCN} and protein structure prediction~\cite{Jumper2021HighlyAP}.

Sharing and using standard datasets accelerate advancements in science and technology~\cite{Barabsi2016NetworkS}. 
Through our random investigation on over 10 papers presenting new network datasets (including their associated websites), we found that they describe network datasets in only 1-2 short paragraphs listing the genre and basic statistics. 
As different network dataset providers make their own 
\begin{figure}[t]
\begin{minipage}[b]{0.4\linewidth}
\includegraphics[width=\linewidth]{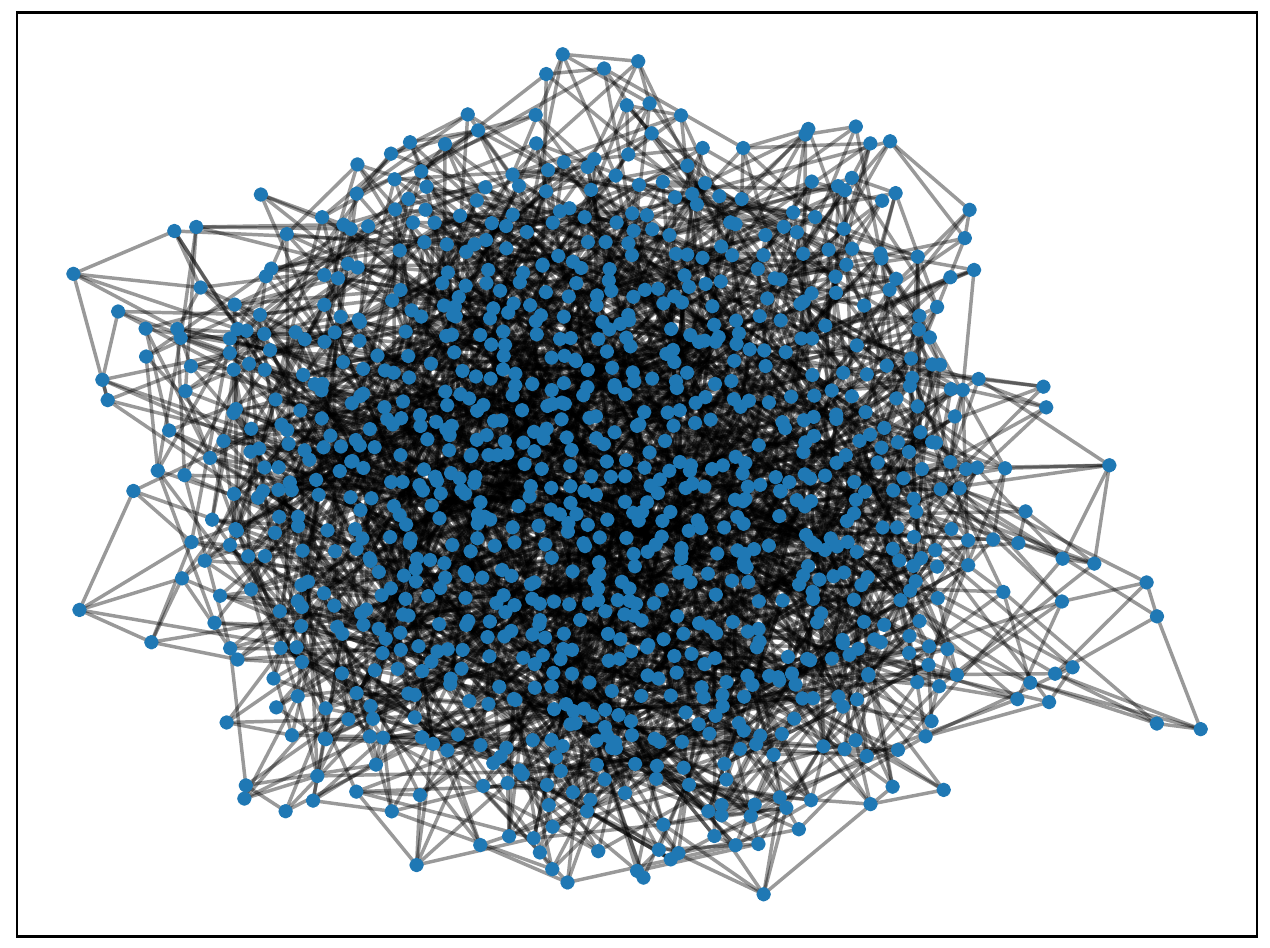}
\caption{Spring positioned visualization for a Watts Strogatz Graph~\cite{Newman1999RenormalizationGA} with 1000 nodes and 3204 edges}
\label{fig:clutter}
\end{minipage}
\quad
\begin{minipage}[b]{0.4\linewidth}
\includegraphics[width=\linewidth]{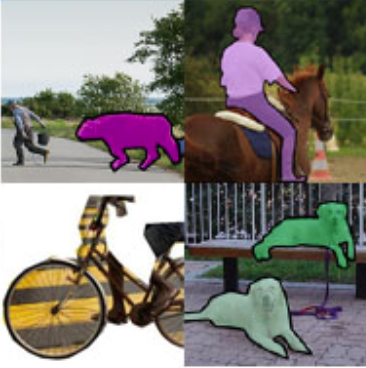}
\caption{Image examples with instance segmentation from Coco dataset on its official website~\cite{Lin2014MicrosoftCC}}
\label{fig:mscoco}
\end{minipage}
\vspace{-0.5cm}
\end{figure}
effort in documenting the datasets, important information is often unintentionally
omitted.
Various data quality issues in common network benchmarks have been pointed out recently~\cite{Toutanova2015RepresentingTF, Hu2020OpenGB}.

In natural language processing (NLP), researchers have found that critical information such as `what language is included' is often omitted~\cite{Bender2011OnAA}.
Proposals have been made and adopted to standardize how to document NLP datasets~\cite{Bender2018DataSF}. 
To address the gap in documenting datasets for machine learning (ML) more broadly, \citet{Gebru2018DatasheetsFD} proposed Datasheets for Datasets, which quickly became the norm for data reporting in industry and academia. 

However, networks are ubiquitous and have unique properties that warrant an extension of these formats.
First, networks are created and researched across different areas and domains (\eg computer science, physics, biology, etc). 
Assumptions and practices are not always shared across these communities, therefore increasing the need for structured communication.
Take for example a social network dataset. 
Social science researchers build the social network through surveys, interviews, etc, and conduct social science research. 
Network science researchers study the community structure. 
ML researchers study how to predict links. 
High performance computing researchers study how to compute social networks efficiently. 
Thus reporting network datasets should generalize to different areas and domains.
Second, networks have complex structures and (often) contain thousands (millions) of nodes, so it is hard to communicate what a network `looks like'. 
Lists and tables are often easier to illustrate. Often one can show a representative instances of \eg a computer vision dataset (images with object annotations as shown in \autoref{fig:mscoco}).
In contrast, for networks, visualizing an instance (entity/node) and its neighbourhood does not reveal the structure of the whole network (sometimes even visualizing the neighbourhood could be hard). 
Another way to communicate  what a network `looks like' is to visualize the whole network.
Unfortunately, visualizing large networks has proven to be difficult (as shown in \autoref{fig:clutter}) ~\cite{Landesberger2011VisualAO}. 
Thus, basic statistics (\eg degree) over networks are used as a tool to communicate network characteristics. 
Moreover, the relational nature of networks make data collection and preparation more complex. There are data preparation operations unique to networks that are not covered by previous proposals for reporting datasets. 
For example, dataset providers can transform a network dataset based on its structure (\eg extracting k-cores).
Some data preparation operations that may be less emphasized in tabular data preparation could have unintended and complex effects on the network structure due to the non-i.i.d. nature of networks (i.e., nodes/samples are not observed i.i.d.). 
For instance, randomly sampling a scale-free network (i.e., degree distributions that roughly follow a power law) could result in a sub-network that is not scale-free~\cite{Stumpf2005SubnetsOS}.
These statistical pitfalls should be documented by network providers.

Current practice in reporting network datasets could result in inappropriate documentation leading to information gaps between network dataset providers and consumers. 
Thus, (unintentional) dataset misuse and misunderstanding could increase, adversely affecting science and society. 
Deployment of applications evaluated or trained with inappropriate networks could have serious consequences such as discrimination against minority groups~\cite{Stoica2018AlgorithmicGC}.

To address the information gap between network dataset providers and consumers, we design \systems as practice for documenting network datasets.
Our development was driven by formative interviews with 12 network practitioners (including faculty, graduate students, research scientists, and engineers in industry). We sought to understand what they would like to see when they encounter a new network dataset. 
A \system is a structured description following a particular template that summarizes and contextualizes a network dataset.
To handle the unique challenges of reporting networks,  we advocate for standard measures (\eg degree) of the network structure, demographics information (if the dataset is related to people), potential bias that already exists or was introduced during data processing, and potential or known errors in the dataset. 
For network dataset providers, we anticipate that \system could relieve the burden of thinking about what to include and how to present the dataset intuitively.
We anticipate that network dataset consumers can assess whether the dataset is appropriate for their task, compare different datasets, and understand limitations.
We hope that \systems  prevents unintentional misuse of network datasets and mitigate bias in high-stakes applications.
While we think that the proposed format of \system covers the most important properties of networks, we see it as a living format that the community continues to refine.

\section{Related Work}
A growing body of literature designs and evaluates structured documents or checklists for datasets, models, and services.   
One of the most prominent works is `Datasheets for Datasets'~\cite{Gebru2018DatasheetsFD}. The authors proposed to accompany documentation with datasets to encourage transparency and accountability for data used by the ML community. 
The framework advocates for systematically documenting  datasets’ purpose, composition, collection process, preprocessing, uses, distribution, and maintenance.
A similar proposal is `Data statements for NLP', which focuses on NLP datasets and addresses ethics, exclusion, and bias issues in NLP datasets and systems~\cite{Bender2018DataSF}.
`Dataset nutrition labels' describes the concept of nutrition label to assess the quality and robustness of a dataset before it is used in model development~\cite{Yang2018ANL, Holland2018TheDN}.
Recent studies have also investigated various aspects of documentation in data science lifecycle such as data readiness~\cite{Beretta2018EthicalAS}, industrial data practices for computer vision datasets~\cite{Miceli2021DocumentingCV}, academic dataset usage~\cite{Geiger2020GarbageIG}, accountability for ML datasets~\cite{Hutchinson2021TowardsAF} etc. 
Beyond datasets, `model cards' were proposed to accompany trained ML models including information like intended use cases, evaluation data, performance measures  etc~\cite{Mitchell2019ModelCF}. The goal of `model cards' is to improve transparency between stakeholders of ML models. 
`Factsheets' and `AI fairness checklists' have a similar motivation, but were targeted at documenting AI services and systems~\cite{Hind2019IncreasingTI, Madaio2020CoDesigningCT}.

Complementing existing work, we primarily focus on networks, and develop a structured way to communicate network datasets.
Our work is functionally similar to `Datasheets' and `Data statements' in the context of ML, but we also consider challenges outside ML as networks are used and shared across different areas and domains.

\section{Preliminaries}
We briefly discuss some terms and symbols in this section. 
``Graph'' is more prevalent in the ML community, but ``network'' has been popular in the data mining and network science communities.
We primarily use ``network'' in this paper, and use ``graph'' for specific terms (\eg knowledge graph). 
The reason is that network analysis is generally concerned with the properties of real-world data,
whereas graph theory is concerned with the theoretical properties of the
mathematical graph abstraction.

A network $G$ is represented as a tuple $G=(V, E, \mathbf{X}^V, \mathbf{X}^E)$, where $V$ is the set of nodes, $E$ is the set of edges, $\mathbf{X}^V$ is the attribute matrix of nodes, $\mathbf{X}^E$ is the attribute matrix of edges. Nodes $V$ represent entities, and edges $E$ represent relationships between entities. $\mathbf{X}^V$ represents node attributes. $\mathbf{X}^E$ represents edge attributes. Each row of $\mathbf{X}^V$ ($\mathbf{X}^E$) represents one node (edge), and each column of  $\mathbf{X}^V$ ($\mathbf{X}^E$) represents one attribute. $\mathbf{D}$ denotes the degree matrix, where the degree of each node is on the diagonal. $\mathbf{A}$, $\mathbf{L}$ denote the adjacency matrix, the Laplacian matrix of the network respectively. We will use this terminology throughout the rest of the paper. 

\input{template}
\section{Development Process}
We started with formative interviews with 12 network practitioners including faculty, graduate students, research scientists, and engineers in industry. We asked interviewees at least: what kind of network data they would use; what kind of graph problems they are (have been) working on, and what action they would take when encountering a new network dataset. 
 
Next, using our experience as network researchers, we analyzed potential use cases of networks in academic and industrial contexts. In academic contexts, the \system should provide readers enough relevant information to understand the academic work. Given an arbitrary network, the best algorithm for an analytical problem (\eg community detection) depends on the characteristics (\eg size, cohesion, spectrum etc.) of the network. 
For instance, community detection based on asymptotical Surprise~\cite{Traag2015DetectingCU} works well for many small communities, while Louvain community  detection~\cite{Blondel2008FastUO} based on modularity favours networks with few large communities.
This is due to the no free-lunch theorem, which states that any two optimization algorithms are equivalent when averaged across all possible problems (datasets)~\cite{Joyce2018ARO}. 
As network algorithms could favor certain types of networks as opposed to others, when reading scientific work, researchers need sufficient information of the network datasets to understand the performance of the proposed methods, and how the proposed methods can or cannot be expected to achieve similar results on other datasets. 
The aforementioned examples highlight the need to include network characteristics measures. 

In industrial contexts, deployment of applications evaluated or trained with inappropriate networks could have serious consequences~\cite{Stoica2018AlgorithmicGC}. 
A recommender system trained and deployed with different demographics, inappropriate time window or shifted customer inclinations could result in bad user experience and revenue lost~\cite{Koren2009CollaborativeFW}.
Recent studies revealed that ML models reproduce or amplify unwanted societal biases reflected in training data~\cite{Buolamwini2018GenderSI}. 
For example, knowledge graph embeddings have been shown to encode that men are more likely to be bankers, and women more likely to be homekeepers~\cite{Fisher2019MeasuringSB}. 
A question answering system empowered by the knowledge graph may discriminate female. 
Adding a standardized documentation can make application developers be aware of the potential consequences when using the data and infer what additional steps are necessary.
These examples highlight the need to document population, bias and fairness (for applications) when sharing a network dataset. 

We prototyped a template for \systems based on what we learned. We then ``tested'' the template by creating the network report for the EU email dataset~\cite{Leskovec2007GraphED}. While creating the \system, we noted problems like non-intuitive orders, lack of clarity, lack of coverage, and redundancies. We then refined our initial template and recreated the above network report. We incorporated feedback from other network practitioners with two cases (\url{https://tinyurl.com/mwwunkae#case}) and finally created the template present in the next section. Finally, we conducted an initial user study in \autoref{sec:study}, and incorporated the feedback from participants in the final version presented here.

\section{Proposed Template}

\autoref{tab:template} shows the skeleton of the proposed template.
We include four sections in \system: (i) \textit{Curation Rationale} contains the metadata and the contexts of the dataset; (ii) \textit{Dataset Collection, Preprocessing and Annotation} contains the data preparation information; (iii) \textit{Uses} contains the usage information; (iv) \textit{Network Statistics} contains the statistical properties of the dataset. 
The proposed sections are not intended to be exhaustive. 
Additional details may include domain-specific information, legal information, etc. 
Network dataset providers are also encouraged to review and integrate information in datasheets~\cite{Gebru2018DatasheetsFD}.
Here we focus on network aspects of documenting a dataset.

\subsection{Curation Rationale}

This section helps network dataset consumers understand the basic metadata of the dataset, and sets the context for the remaining sections. After reading this section in the report, readers should have a good overview of the dataset. 
\\
\textbf{Author(s) and reference of the network dataset}: 
This acknowledges the authors who share the network. If the network dataset in an academic paper, it is encouraged to list the reference.
\\
\textbf{Purpose}: Why was the network dataset created? Was there a specific usage in mind? Network dataset providers typically create the dataset primarily for a specific use case. 
This point could inform users of the reasons for certain decisions (such as collection, preprocessing and annotation). 
\\
\textbf{Domain}: Which domain does the network lie in? 
Domains include online social networks, metabolic networks, citation networks, etc.
\\
\textbf{Node and edge semantics}:
What does each node and each edge mean? For instance, in a Twitter who-follows-who network, a node is a user, while an edge is the following relationship.  Node and edge semantics form the cornerstone for all tasks (\eg link prediction, node classification, etc.). 
\\
\textbf{Contents}: 
This includes a brief plain text description on dataset contents. Indicating if the network is a snapshot sets the context for network sampling (\autoref{subsubsec:collect}), and affects tasks like parameter estimation and simulation~\cite{Leskovec2006SamplingFL, Ahmed2013NetworkSF}.
\\
\textbf{Types of network(s)}:
Many tools and algorithms expect certain types of networks, so it is necessary to provide a precise description of the network types.
Networks can be categorized based on node properties, edge properties and hybrid properties.
We provide a taxonomy of network types in \autoref{tab:template}, where each entry is a possible tag for the network type. 

\subsection{Dataset Collection, Preprocessing and Annotation}
This section explains how the author(s) created the network dataset. 
After reading this section, network dataset consumers should know whether the dataset is compatible with their tasks.
For example, if the provider removed the timestamp of a Twitter who-follows-who dataset, then this dataset is not suitable for modeling time-evolving behavior.
Through our investigation and interviews with network practitioners on network dataset creation, we present frequent operations on network data preparation, and emphasize how certain operations on a network dataset could have unintended impacts on different tasks. 
We also advocate for including demographic information in this section if the network is related to people or the annotation process is conducted by people.

\subsubsection{Data collection} \label{subsubsec:collect}
How was the data collected? Was there any known error or bias introduced in the data collection process? 
The ideal collected data would be a census or enumeration of the network (\eg a friendship network within a 50-people company), which includes every node, and every edge between nodes. Unfortunately, this is not feasible under most circumstances. As a result, a typical collection includes sampling, which refers to selecting a sampled network that can represent the population. Since network sampling is often part of data collection, we  explicitly include network sampling in the data collection. We refer to `raw data' the collected data prior to any processing.

\textbf{Data collection mechanism and raw data description}:  What was the data collection mechanism? Was there any error (\eg missing nodes, erroneously omitted edges) or bias (\eg bias to high-degree nodes) introduced by the data collection mechanism?
The data collection mechanism determines the format, the semantics, and the quality of the raw data, thus affecting all downstream data processing choices. 
In some domains, nodes or/and edges do not naturally exist in the raw data, and network dataset providers construct networks from the raw data. 
For instance, some brain networks are constructed from multiple time series signals, where each time series represents the activity for one region in the brain~\cite{Rubinov2010ComplexNM}.
However, in tabular data, the meaning of each instance (sample) does not change (\eg an image collected from the web is always an image in the dataset).
Dataset providers should describe the raw data. 

The data collection mechanism is a significant source of error when making measurements over network structures.
In biology, metabolic, protein, and regulatory networks are typically measured in a lab environment. 
Suffering from natural variation in biological systems and inconsistent measurement conditions, experiments usually don’t give the same results every time, meaning that any individual measurement may be an error~\cite{newman2018networks}. 
Social networks collected from surveys or questionnaires could suffer from missing data, where participants may decline to answer some or all questions~\cite{Marsden1990NETWORKDA, Bernard1977INFORMANTAI}. 
The error or/and bias introduced in the data collection should be included in \systems, thus network dataset consumers could infer whether the error or/and bias would influence their tasks.

\textbf{Network sampling}:
 Was there any sampling process when collecting the network dataset? What sampling strategy has been chosen? What was the reason for the sampling strategy? 
Network sampling often inherently happens when the data is collected. 
For example, researchers use breadth first search to  crawl online social networks automatically~\cite{Mislove2007MeasurementAA}.
Many factors make it difficult to study networks in their entirety.
Some networks are not completely visible to the public (\eg Facebook), or can be accessed only through crawling (\eg the Web). 
In other cases, the measurements required to observe the underlying networks are costly (\eg experiments in biological network). 
Moreover, many networks are continuously changing (\eg telecommunication interactions, user-product interactions) and can never recorded in their entirety.For these networks, providers should describe the collection time frames.

The non-i.i.d. nature of networks makes network sampling complex. 
For example, given a population network, randomly sampling a set of nodes versus randomly sampling a set of edges would result in very different network structures. 
Network sampling could introduce sampling bias to the network, which makes it hard to correctly estimate the parameters of the whole network.
For instance, sampling in a breadth-first manner is biased as high‐degree nodes have a higher chance of being sampled~\cite{Kurant2010OnTB}.
People who intend to study hard-to-reach communities should not use breadth-first sampled social networks.

With the detailed sampling information, network dataset consumers could examine whether the network dataset is biased towards particular properties and whether it is still suitable for their task, and what additional processing steps are needed. For instance, researchers have used Metropolis-Hasting random walk to induce an unbiased sample of Facebook users~\cite{Gjoka2010WalkingIF}. 

\subsubsection{Data preprocessing}
How was the network dataset generated from raw data? Preprocessing converts raw data to a structured network format. Steps may appear in different orders or can be repeated~\cite{Hameed2020DataPA}. 
Here we discuss six typical steps to preprocess a network dataset: network construction, data cleaning, data filtering, network transformation, attribute transformation, and data split. 
Data split is common for ML tasks, while the others are common in data preparation.
There is no strict order for these steps, thus network dataset providers can interchange them under different circumstances.
 If network dataset providers provide the raw data, the data preprocessing part in \system allows network dataset consumers to replicate the steps.

\textbf{Network construction}: What strategies were used to transform raw data to nodes and edges? What are the reasons for choosing these strategies?
When the raw data does not naturally include nodes or/and edges, network construction is a crucial step to set up nodes or/and edges.
Essentially, network construction is to set up $V$ and $E$ in the network.
This step is unique to network datasets as tabular data does not have relations between instances.
Network construction determines the semantics of nodes or/and edges, so it should be consistent with `node and edge semantics' in `Curation Rationale'.
Different network construction methods fundamentally change the network structure and the compatible tasks.
For example, in brain networks, the edges could represent anatomical tracts or functional associations, and the semantics largely determines the neurobiological interpretation of network topology~\cite{Rubinov2010ComplexNM}. 

\textbf{Data cleaning}: How were nodes, edges, or networks (in a multi-network dataset) edited or augmented? 
Data cleaning refers to the removal, addition, or replacement of less accurate (or inaccurate) data values with more suitable, accurate, or representative values~\cite{Hameed2020DataPA}. 
Research on data cleaning for networks has shown that inappropriate data cleaning strategies could lead to a high error rate in wireless sensor networks due to power exhaustion and environmental noise~\cite{Cheng2018DataQA}. 
Thus, including data cleaning details in a \system is necessary for network dataset consumers to examine if the data quality is appropriate for the tasks and if they need to conduct additional data cleaning steps. 
Typical operations of data cleaning on networks include: removal of nodes, removal of edges, removal of attributes, node deduplication, etc.  
Data cleaning has a direct impact on data quality and data usage. 
Removal of specific attributes like timestamps could limit the applicability of the network dataset.

\textbf{Data filtering}: What nodes or/and edges have been filtered out? What was the reason for filtering?
Data filtering refers to generating a subset of nodes or/and edges~\cite{Hameed2020DataPA}. 
Data cleaning focuses on removing inaccurate values, while data filtering focuses  on simplifying the contents of the data.
Different from tabular data, where data filtering is done based on attribute value(s) (value filtering), in networks, data filtering could be based on the network structure (structural filtering). 
Value filtering is to simplify contents based on $\mathbf{X}^V$ and $\mathbf{X}^E$, while structural filtering is based on $V$ and $E$.
Structural data filtering could be a potential source of bias in the network.
For instance, extracting k-cores of a large network, a common method~\cite{Safavi2020CoDExAC, Ni2019JustifyingRU}, biases to high-degree nodes.
Both value filter (on nodes or/and edges) and structural filter should be detailed in this part. 

\textbf{Network transformation}: How was the network transformed? 
Network transformation refers to converting the network from one format  to another. 
Typical operations include the projection of a bipartite network (\eg constructing a user similarity network based on the number of co-purchased items), the projection of a multilayer network, and aggregation of nodes for hypergraphs. 
For projection, we encourage network dataset providers to carefully think about the weighting scheme, because the redistribution of weights has a strong effect on the community structure (especially in dense networks)~\cite{Fan2007TheEO}.

\textbf{Attribute transformation}:
How were the node or/and edge attributes transformed? Network transformation is based on network structure, while attribute transformation focuses on attributes of nodes or/and edges ($\mathbf{X}^V$ and $\mathbf{X}^E$). 
Typical transformations include value normalization, value scaling, dimension reduction, change of attribute names, etc. 

\textbf{Data split}: How was the data split to train/validation/test sets? What was the splitting ratio?
ML researchers use data splits to train and evaluate their models.  
Tabular datasets are often split to train/validation/test sets by random selection. 
However, randomly selecting nodes and edges for training and testing leads to overly optimistic accuracy due to the non-i.i.d. nature of graphs~\cite{Lohr1999SamplingDA}.

One common mistake in data splits is unintentional data leakage. Leakage refers to the use of information in the model training process that would not be expected to be available at prediction time, causing the predictive scores (metrics) to overestimate the model's utility when run in a production environment~\cite{Kaufman2011LeakageID}.
We want to highlight it in data splits because network related tasks often train and test on a single network, thus having a high probability of leakage.
Researchers have found that leakage existed in a common knowledge graph completion (link prediction) benchmark~\cite{Toutanova2015RepresentingTF}, where the relation set of inverse entity pairs is almost the same \eg comparing entity pairs $[e1, e2]$ for relation $r$ to the entity pairs $[e2, e1]$ for relation $r^{'}$ (reverse $r$). 
Thus including relation $r$ between entity pairs $[e1, e2]$ in the training set leaks information because in real use cases, $r$ would not exist in the knowledge graph. 
As a result, prior models were not fairly evaluated. 
The main reason for this mistake is that the links between entities were split randomly into train/validate/test set. 

Besides leakage, the notion of `transductive learning' and `inductive learning' brings additional ambiguity when evaluating ML models on graphs and networks. 
In the context of graph machine learning, `transductive learning' refers to seeing the whole graph (including the test set but without the labels of the test set) beforehand, while `inductive learning' assumes there are unseen nodes added to the graph during test time. 
In an inductive node classification task, researchers have found duplicated nodes (near 5\%), edges and labels in training and testing sets~\cite{Hu2020OpenGB} due to data splitting. 
This violates the inductive assumption as the duplicated nodes in test sets cannot be viewed as newly added. 
To avoid similar issues, we encourage network dataset providers to carefully examine the data splitting strategy and provide details on how the strategies are implemented. 
We also encourage network dataset consumers, especially ML researchers, to check if the data is split appropriately for their tasks.

\subsubsection{Instance Demographics}
In case the network dataset is related to people, what are the demographic characteristics of the instances? 
Possible demographic information include but not limited to age, gender, race/ethnicity, and socioeconomic status. 
Including instance demographics is important for ML applications in real-world deployments.
Researchers found that models can reproduce or amplify unwanted societal biases reflected in training data~\cite{Buolamwini2018GenderSI}.
In real-world deployments, biased data can result in unfair decisions and discrimination of minority groups, which can lead to severe consequences in high-stakes domains. 
For instance, recommender systems that make predictions based on observed data can easily inherit bias that may already exist~\cite{Yao2017BeyondPF}.
Knowledge graph embeddings have been shown to encode that men are more likely to be bankers, and women are more likely to be homekeepers~\cite{Fisher2019MeasuringSB}.
To mitigate bias and improve fairness, sensitive attributes (\eg age in a loan and debt network; gender in a job posting network~\cite{Blommaert2020TheGG, Bose2019CompositionalFC}) involving discrimination should be available to network dataset consumers. 

\subsubsection{Data annotation}
Who annotated the data (\eg crowds, domain experts)? How was the data annotated? Was there a shareable user interface? Were the labels noisy or clean? How are labels derived from multiple noisy labels? 
The description of data annotation process is detailed in this part. Including details at the data annotation process in \systems is useful for network dataset consumers who would like to use the labels. 

\subsubsection{Annotator Demographics}
What are the demographic characteristics of the annotators? 
Possible demographic information include but not limited to age, gender, race/ethnicity, socioeconomic status, native language, and expert level. 
In social networks measured using surveys or questionnaire, subjectivity is a notable source of error when measuring network structure~\cite{newman2018networks}, especially when respondents give ambiguous answers and some interpretation is required to decide what those answers mean. 
Annotators' native language could impact converting qualitative answers to quantitative metrics. 
To curate a knowledge graph of political data, annotators' demographics information could influence their political stands, thus resulting in disputable annotations. 
Expertise level matters for domain-specific network annotations. 
For instance, domain experts need to perform biology experiments to annotate whether a protein-protein interaction network is an enzyme. 
The annotator demographics could help network dataset consumers assess the quality of the annotations and potential sources of errors. 

\subsection{Uses}
This section  allows network dataset consumers to quickly grasp what the dataset
should and should not be used for. We include three parts: primary use and  other uses. We understand that the use space of the network dataset could be huge, and it is impossible to list every use case. We encourage network dataset providers to list typical ones in this section. 
\\
\textbf{Primary intended uses}: 
This part details the use cases and tasks that the network dataset providers have in mind when creating the dataset. 
For machine learning datasets, this part should include the task type (\eg classification, regression), task category (\eg node, link or graph) and the evaluation metric(s) for the task.
When there are no particular tasks for network dataset providers (\eg A social network company releases part of the user network data to the public), the primary intended uses could be broad.
\\
\textbf{Other uses}:
This part details the potential uses cases of the network dataset. Examples include offline anomaly detection on a streaming network dataset, benchmarking HPC platforms on an online social network dataset, etc. 
\subsection{Network Statistics}
This section includes point statistics and distributions of the network(s).
The complex structure of networks makes it hard for people to understand whether the dataset is appropriate for their tasks by eyeballing the plain text description, a small subset (\eg the egonet of one node), or one network plot. 
Network statistics could partially reveal such complexity with a few numerical values and visualizations of a few distributions. 
This section is inspired by the electronics industry, where standard  characteristics (\eg operating temperature) shown by numerical values and visualization of essential properties (\eg time vs. frequency) are available in datasheets~\cite{miniatureAluminum}. 
The electronics industry set such standards because small deviations that may seem insignificant can have severe consequences for the system as a whole~\cite{Gebru2018DatasheetsFD}. 
However, the generative process of networks is even more complex, as the same configuration in graph models could result in significantly different networks. Similar to the datasheet in the electronics industry, we propose that a \system includes point statistics and visualizations. 
We list point statistics in \autoref{subsubsec:point}, where each measure is a numerical value. We list distributions in \autoref{subsec:distviz}, where visualization of distributions are shown.

Given the rich literature in network science, numerous statistical properties could be included in this section.
We design the list through our formative interview, runtime concerns and diversity concerns. 
We prefer fundamental, well-studied measures over complex, domain-specific ones. 
Practically, network datasets could be massive (\eg a billion nodes). 
Thus computationally heavy properties (yet important) like betweenness centrality and diameter are not included in the list, as both of the properties involve calculating the shortest paths of all pairs of nodes. 
We also prefer statistical properties that can cover diverse aspects of networks. 
Thus we choose point statistics and distributions that can cover (i) node and edge characteristics (size, degree, attributes); (ii) spectral properties (spectral radius, algebraic connectivity, singular value distribution, pagerank distribution); and  (iii) network cohesion (connected components, average clustering coefficient, clustering coefficient distribution, degree assortativity coefficients, max k-core).
Those point statistics and distributions could reveal fundamental phenomenons of the underlying graph (\eg the scale-free property, the small-world phenomenon). 

Visualizations help network dataset consumers to understand distributions.
Patterns in the visualization reveal the dynamics of the network.
Abnormal patterns in the visualization (\eg a spike) could help network dataset consumers understand the networks' anomalies and further infer whether the anomalies would influence their tasks.
We discuss common practices of visualizing network statistics in \autoref{subsec:distviz}.


\subsubsection{Point statistics}:\label{subsubsec:point}
Point statistics map a network’s structure to a simple numerical space. For a multi-network dataset, network dataset providers can provide the average (and standard deviation) point statistics of all the networks in the dataset. 
Other point statistics beyond our list are also encouraged (\eg structural balance for signed networks).

\textbf{Number of nodes ($|V|$) and number of edges ($|E|$)} describe the size of the network. Network dataset consumers can roughly understand how much computation resources are necessary for processing the network.

\textbf{Proportion of nodes in the largest connected component} reveals the basic network connectivity. For directed networks, both weakly connected and strongly connected components should be considered. 

\textbf{Average degree ($\bar{d}$)} is defined as
 $\bar{d} = \frac{1}{|V|} \sum\limits_{u\in V} deg(u)$, 
where the function $deg(\cdot)$ denotes a degree measure (\eg in degree) of a node. 
For undirected networks, $\bar{d} = \frac{2|E|}{|V|}$. 
For directed networks, the average out degree equals the average in degree, and both can be calculated as $\bar{d_{out}} = \bar{d_{in}} = \frac{|E|}{|V|}$
It quantifies the extent to which a node is connected to other vertices within the network.

\textbf{(Tail) power-law exponent ($\gamma$)} characterizes the degree distribution of the network. Many networks follow a degree distribution power law, \ie the number of nodes with degree $n$ is proportional to the power $n^{-\gamma}$, for a constant $\gamma$ larger than one. There are multiple ways of estimating $\gamma$, we recommend estimating $\gamma$ using the robust method defined as
$ \gamma = 1 + |V|\left(\sum\limits_{u\in V} \ln\frac{deg(u)}{d_{min}}\right)^{-1}, $
where $d_{min}$ denotes the minimal degree~\cite{newman2018networks}.
For networks that do not follow a power-law degree distribution, network dataset providers can indicate N/A for this item. N/A also informs network dataset consumers that the dataset is not scale-free.

\textbf{Spectral radius ($\rho$)} and \textbf{Algebraic connectivity ($a$)} reveal mathematical properties on graph spectrum. Let $\lambda_1,\cdots,\lambda_n$ be the (real or complex) eigenvalues of $\mathbf{A}$. Then
$ \rho = \max(|\lambda_1|,\cdots,|\lambda_n|)$. 
Algebraic connectivity $a$ is the second smallest eigenvalue of $\mathbf{L}$.

\textbf{Average triangle count ($t$)} counts how many triangles one node is involved in. Triangles have been used in spam detection, link recommendation, and graph random graph models, etc~\cite{Tsourakakis2010CountingTI}. We recommend to calculate it as $ t = \frac{|\{ (u, v, w)\vert \text{$u,v,w$ are connected} \}|}{6}$, because it is independent of the orientation of edges when the network is directed. Here

\textbf{Average clustering coefficient ($c$)} quantiﬁes the extent to which pairs of nodes with a common neighbor are also themselves neighbors. It extends the triangle count of a node by normalizing the triangle count with the number of pairs of nodes in its neighbourhood. It is used as an indicator of ``structural holes'' in a network, which has rich applications in social networks~\cite{newman2018networks}. As betweenness centrality (BC) is computationally intensive, average clustering coefficient can also be thought of as akin. Betweenness measures a node’s control over information ﬂowing between all pairs of nodes in its component, while clustering clustering is a local version of betweenness that measures control over ﬂows between just a node’s immediate neighbors~\cite{newman2018networks}. Mathematically, it can be calculated as $ c = \frac{3t}{\sum\limits_{u\in V} \frac{1}{2} deg(u) (deg(u) - 1)}$.

\textbf{Degree assortativity coefficient ($r$)} quantifies the extent to which connected nodes have similar degrees. It is the Pearson correlation between the degree of connected nodes. Mathematically, it can be calculated as 
$ r = \frac{\sum_{(u,v)\in E} (deg(u) - \bar{d})(deg(v) - \bar{d})}{\sum_{(u,v)\in E} (deg(u) - \bar{d})^2}$.
Besides, network dataset providers can also provide assortativity  mixing by other measures of interests (\eg Pearson correlation between the attribute of connected nodes). Some technologies related to networks such as graph convolution neural networks would fail if the network is not near homophily (\ie connected nodes are similar)~\cite{Kipf2017SemiSupervisedCW}, thus providing one measure of homophily could help prevent misuse of those network technologies.

\textbf{Max $k$-core ($k$)} is the maximum $k$ such that $k-$core exists in the network, and a $k-$core is the maximal subgraph where all vertices have degree at least $k$. $k-$core has been applied to diverse areas such as hierarchical structure analysis, network visualization and network clustering~\cite{Shin2016CoreScopeGM}. Max  $k-$core also provides the upper bound for max clique (\ie another important structure in networks).

\subsubsection{Distributions of node/edge statistics and attributes}\label{subsec:distviz}
We now discuss the distributions in \system and common design choices for visualization.
For a multi-network dataset, network dataset providers can treat the dataset as a whole network, where each component is one network inside the dataset. 

\textbf{Degree distribution ($deg(\cdot)$)}:
The degree distribution is typically a skewed distribution, as there are more small-degree nodes than large-degree nodes.
If the tail degree distribution follows a power law, then it's typical to plot degree distributions on a log-log scale (\ie both axes have a logarithmic scale)~\cite{newman2018networks}.
Histograms, line charts and point charts are common encoding choices. 
Histogram involves binning, and inappropriate binning can hide the trend in the data.
For simplicity, we recommend point charts or line charts with degree on the x-axis and frequency (or probability mass function, or inverse cumulative distribution function \footnote{Let $X$ be a random variable with cumulative distribution function (CDF) $F$. The \textit{inverse} CDF is defined as $F^{-1}(q) = \inf\{x: F(x) > q\}$~\cite{Wasserman2004AllOS}.}) on the y-axis.

\textbf{Pagerank distribution ($pr(\cdot)$)}~\cite{Page1999ThePC}:
Pagerank distribution can be thought of as an extended form of degree distribution, which considers how many neighbors a node has and how important those neighbors are. It can be calculated as 
$ pr(G) = (\mathbf{I}-\alpha \mathbf{AD^{-1}})^{-1}\mathbf{1}, $
where $\alpha = 0.85$.
Given its similarity to degree distribution, similar design choices in axes (\eg log-log scale for scale-free networks) could be used.
One significant difference is that pagerank is a continuous variable, while degree is a discrete variable, where the value is an integer at most the max degree. 
So we recommend point charts or line charts with pagerank value on the x-axis and inverse cumulative distribution function on the y-axis.

\textbf{Singular value  ($\mathbf{\Sigma}$) (or eigenvalue ($\mathbf{\Lambda}$)) distribution}:
Singular values (and eigenvalues) reveal the spectral properties of the graph. Spectral graph properties have broad applications in graph partitioning (clustering, community detection), dynamical systems~\cite{newman2018networks}. 
There are variations of graph matrices ($\mathbf{A}$, $\mathbf{L}$, $\cdots$) and spectral decomposition (eigenvalue decomposition, singular value decomposition, $\cdots$ ) that characterize spectral properties.
One common way to reveal the graph spectrum is to use the eigenvalues of the adjacency matrix~\cite{newman2018networks}.
However, for directed graphs, the eigenvalues are complex numbers, which brings additional difficulty in visualization.
We recommend using the singular values of the adjacency matrix, which bypasses the complex number problem, and gives a unified view on both undirected and directed networks. 
Calculating the exact solution of all singular values is computationally expensive, so we recommend computing the approximate top-k singular values.
The distribution can be plotted as a line chart or a point chart can be the encoding choices.

\textbf{Clustering coefficient distribution ($CC(\cdot)$)}:
Clustering coefficient (\ie local clustering coefficient) is the fraction of pairs of neighbors of node that are themselves neighbors. $\forall u\in V$,
$ CC(u) = \frac{\text{number of pairs of neighbors of $u$ that are connected}}{\text{number of pairs of neighbors of $u$}}.$
Clustering coefficient distribution is closely related to the small-world phenomenon of real-world networks~\cite{Watts1998CollectiveDO}.
It is a continuous variable with values between 0 and 1. 
A common practice to visualize clustering coefficients as a cumulative distribution function on points or lines with linear scales~\cite{newman2018networks}.

\textbf{Node attributes}:
For a quantitative attribute, a common practice is to use histograms. 
For an ordered attribute, a bar chart can be used to show the frequency of each category.
For a categorical attribute, network dataset providers can use a bar chart to indicate the frequency of each category, and arrange the categories according to the frequency.
Horizontal bar charts make it easier to have readable labels than vertical bar charts. 
If there are too many categories, one typical practice is to list top-k categories in the bar charts, and use `others` to indicate the remaining categories. 

\textbf{Edge attributes}:
Visualization of an edge attribute is similar to that of a node attribute for a  quantitative/ordered/categorical one. 
For temporal networks, where each edge contains a timestamp, we commend a line plot with the relative time stamp on x-axis and number of edges within a fixed time window on the y-axis.

\section{Implementation Consideration}
The first and primary consideration for implementing \system is the time cost. 
Documentation could be a burden~\cite{Miceli2021DocumentingCV}. 
We analyze the time cost of the \system in terms of objective information (Network Statistics) and subjective information (Curation Rationale; Dataset Collection, Preprocessing and Annotation; Uses). 
The contents of subjective information require input from people, while the objective information could be automatically created.
For large organizations, standard internal software could be adopted to automatically create the objective information. The software could also be tailored for the organization's design style in visualization, graph storage infrastructure, network analysis tools, etc. If there is a centralized platform to submit datasets, certain checks could be automated when submitting the datasets, similarly to CI checks on software repositories. 
For small organizations or individuals, spending a significant time developing standard software could be overwhelming. 
Network dataset providers could write code for the dataset in an ad-hoc manner. 
Nevertheless, if the dataset would update frequently, the code could be reused over time. 
As researchers in the network field, one of the authors spent around 2.5 hours creating the objective information in an ad-hoc manner and around 1 hour creating the subjective information.
When the author started to create more network reports for other datasets, she improved the code to maximize code reuse. 
We make our code publicly available for reference (https://tinyurl.com/mwwunkae).
Compared with the time overhead in data collection and processing, we believe this time for documentation is appropriate. 

The second concern is information availability. 
A lot of network datasets are not created `from the ground up'~\cite{Gebru2018DatasheetsFD} (such as those in TUDataset~\cite{Morris2020TUDatasetAC}), so it is hard to gather all the necessary information. 
In this case, network dataset providers should explicitly indicate \textit{Unknown} (as opposed to skipping it). 
Besides, some of the demographic information, such as gender, is sensitive.
Following a similar concern in model cards, we encourage network dataset providers to use self-identified labels or using labels clearly designated as \textit{perceived} (as opposed to self-identified)~\cite{Mitchell2019ModelCF}.
Looking forward, efforts aimed at collecting fined-grained information and annotation should serve as a key step to strengthen fairness, accountability, and transparency in network technology. 

The third concern is how to encourage implementing \systems. 
Publication venues related to networks could encourage the community to release a \system alongside new datasets. 
Excitedly, ``Datasets and Benchmarks Track'' in NeurIPS 2021 enforces dataset documentation and intended uses accompanying the main paper, and recommends authors to follow existing frameworks such as Datasheets~\cite{Gebru2018DatasheetsFD}, Data Statements for NLP~\cite{Bender2018DataSF}, and Nutrition Label~\cite{Holland2018TheDN}. 
Microsoft, Google, and IBM are adopting similar reports for general AI datasets~\cite{Gebru2018DatasheetsFD}.
Besides, funding agencies could require \systems in data management plans. 
We believe that the \system is a good complement for the above efforts in reporting datasets. 
Ultimately, implementing \system can dramatically improve the utility of the datasets for others, and benefit both direct stakeholders (\eg application developers, funding agencies) and indirect stakeholders (\eg users of recommender systems).

\section{Initial User Study}
\label{sec:study}
\subsection{Study Design}
We conducted an initial qualitative user study to explore the functionality of the \system in an academic environment. 
We hypothesized that \systems facilitate dataset communication and paper understanding.

\textbf{Datasets and Paper}\, We used the paper ``Higher-Order Label Homogeneity and Spreading in Graphs'', and the POKEC dataset  ~\cite{Eswaran2020HigherOrderLH}. This version of POKEC dataset was built on top of the original POKEC dataset present in~\cite{Takac2012DATAAI}. This mimics a common academic scenario, where datasets augment other datasets.

\begin{figure}[th]
\includegraphics[width=0.47\textwidth]{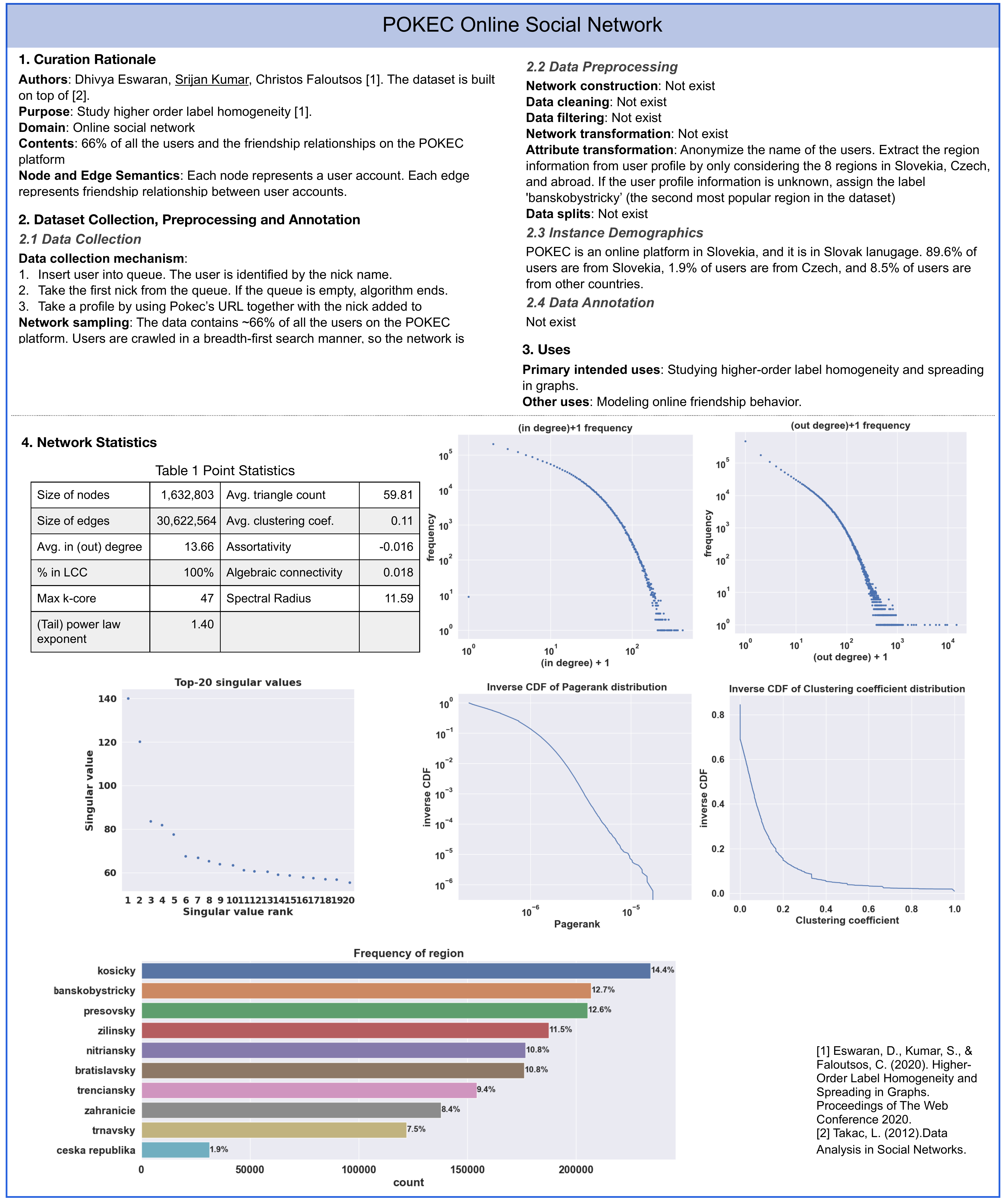}
  \caption{
  Network report for POKEC network (used in user study).}
  \label{fig:pokec}
\end{figure}

\textbf{Study Protocol}\, Each session began with a 15-minute tutorial on the main idea of the paper. We then asked participants to read the dataset description, followed by the network report in~\autoref{fig:pokec}. After that, we asked them to read through the experimental result section in the paper. During the study, we answered questions regarding the paper, but we did not answer questions regarding the dataset. Finally, we did an semi-structured checkout interview. Audio was recorded via Zoom.

\textbf{Participants}\, We recruited 6 participants (3 females, 3 males), all graduate students (1 MS, 5 PhD), with prior ML experience. 4/6 participants have research experience in networks. 2/6 participants have industrial experience related to network data. One of the participants had a hard time understanding the paper, so we discarded this session in analysis.

\subsection{Analysis \& Results}
After transcribing the interviews, two of the authors did a line-by-line open coding on one transcript and produced a coding guide. Then the two authors coded separately on all transcripts, and discussed the codes to produce the results. 
\subsubsection{Gain deeper understanding of academic papers}
We found that all participants have expressed difficulty in understanding network datasets under academic  scenarios. P3 mentioned that \emph{``it’s hard to mentally visualize the dataset by just reading the description in the paper''}. 
4/5 participants mentioned that the \system helps them verify assumptions of a given paper on the dataset. During the study, all participants have attempted to explain the experimental phenomenon with the information in the \system. Participants either confirmed that the dataset aligns with the paper's assumption, or started to critique the assumption. P3 said that  \emph{``All [the network report] helps is explaining why this algorithm may succeed because it gives you a visual description of the dataset''}. We also found that providing the network report with the paper could drive people to think critically about the data and the paper. P4 said that 
\emph{``If there are some ways for me to look at the stats, and how they differ from what I normally expect, probably I would look deeper into the dataset to see if there are some issues in the data generation pipeline.''}
P4 subsequently mentioned that he would not check the assumptions unless he had the report. 

\subsubsection{Build trust in academic papers} 3/5 participants mentioned trust in academic papers in the network domain. P3 said that \emph{``if a paper is claiming things, then I would like to look at some statistics around the claimed things, which are not in the paper''}. P5 said that \emph{``people don't want to explain [datasets] more, or don't want to waste paper real estate to speak about basic stats, like the network report, so I have hard time in paper understanding''}. 2/5 participants mentioned that documentation like \systems are important for building trust especially when the data is not publicly available. 

\subsubsection{Create a structured channel to communicate network datasets}
All participants mentioned that the network report helps people communicate network datasets. P5 emphasized that it creates transparency, and said that \emph{``There are a thousand ways that I could construct a graph from [...] data. Whatever format I am deciding on, I should explain what I have done in detail and the rationale for why I have done it. Otherwise, it seems not genuine enough. And I think it is a great step towards more transparency in the network domain, which is needed.
''} 3/5 participants said that the network report allows for comparison and using data from other domains.  4/5 participants mentioned that the \system saves researchers' time. 
P3 said that \emph{``Section 3 (Dataset Collection, Preprocessing and Annotation) is useful to me. ... I didn't think of [network sampling and data filtering] until I saw it.''} P1 said that \emph{``[The network report] saves time. ...  And if you have a bunch of [network datasets], and you are deciding which dataset to use for your own research. And you can go to the uses section, and the data collection section, and  find out which one is most specific to me.
''}. 2/5 participants mentioned that they would open a Jupyter notebook to make some visualizations, so the section of Network Statistics is very useful. 
As participants find almost all sections of the \system useful, we conclude that the \system creates a structured way to communicate network datasets. 

\subsection{Limitations}
The qualitative study was conducted in an academic environment, so the above results and analysis may bias towards academic use cases. As the user study is an initial study to explore the functionality and usability of the \system, it is not intended to be a complete evaluation of the \system.  

\section{Discussion and Conclusion}
We have proposed the \system as a structured template to report information about network datasets. The \system includes network characteristics, information that captures population, bias and fairness, and detailed data preparation information. 
Our qualitative study suggests that the \system could help facilitate structured communication between network dataset providers and consumers.

Although \systems are designed to be flexible across different domains and use cases, the usefulness of \systems still relies on the integrity of the creators. 
In the near term, it is unlikely that the full contents of \systems could be automatically created and standardized to prevent all inappropriate use. 
We envision a standardized export formats and data loaders for networks accompanying \systems in the future.
Besides, \systems are just one approach to increase transparency in network related technologies, which could include, for example, model cards for pre-trained graph neural network models~\cite{Mitchell2019ModelCF} and vulnerability and robustness testing~\cite{Freitas2021GraphVA}.
Future work includes accelerating the creation of \systems, studying how the information in \systems is interpreted and used by different network dataset consumers: researchers, developers, policy makers, etc, and how \systems could be integrated with other transparent tools. 
\bibliographystyle{ACM-Reference-Format}
\bibliography{sample-base}
\newpage

\end{document}

%% file: template.tex
\begin{figure*}
\begin{mdframed}
\raggedright
\begin{multicols}{2}
    \normalsize{\textbf{Curation Rationale}}\footnotesize
    \begin{itemize}\setlength{\itemindent}{0pt}
    \item Author(s) and reference of the network dataset
    \item Purpose
    \item Domain
    \item Node and edge semantics
    \item Contents
        \begin{itemize}\setlength{\itemindent}{0pt}
            \item Is this dataset a snapshot of a larger network?
        \end{itemize}
    \item Types of network(s)\\
    \begin{itemize}[label={}]\setlength{\itemindent}{0pt}
        \item Based on edge properties:
        \begin{itemize}[label=-]\setlength{\itemindent}{0pt}
        \item directed or undirected network
        \item simple or multi-graph where an edge can repeat
        \item weighted or unweighted network
        \item signed network
        \item temporal network
    \end{itemize}
    \item Based on node properties:
        \begin{itemize}[label=-]\setlength{\itemindent}{0pt}
        \item homogeneous or heterogeneous 
    \end{itemize}
    \item Based on hybrid (node and edge) properties:
    \begin{itemize}[label=-]\setlength{\itemindent}{0pt}
        \item knowledge graph
        \item bipartite network
        \item multilayer network
    \end{itemize}
    \end{itemize}
\end{itemize}
     \normalsize{\textbf{Dataset Collection, Preprocessing and Annotation}}\footnotesize
  \begin{itemize}\setlength{\itemindent}{0pt}
    \item Data collection
    \begin{itemize}\setlength{\itemindent}{0pt}
        \item Data collection mechanism and raw data description
        \item Network sampling 
    \end{itemize}
    \item Data preprocessing 
        \begin{itemize}\setlength{\itemindent}{0pt}
        \item Network construction \textcolor{gray}{(\eg parcellation of brain scans)}
        \item Data cleaning \textcolor{gray}{(\eg removing inaccurate nodes)}
        \item Data filtering \textcolor{gray}{(\eg extracting k-cores)}
        \item Network transformation \textcolor{gray}{(\eg bipartite graph projection)}
        \item Attribute transformation \textcolor{gray}{(\eg anonymize attribute values)}
        \item Data splits
    \end{itemize}
    \item Instance demographics
    \item Data annotation
    \item Annotator demographics
\end{itemize}
    \normalsize{\textbf{Uses}} \footnotesize
\begin{itemize}\setlength{\itemindent}{0pt}
    \item Primary intended uses
    \item Other uses
\end{itemize}
    \normalsize{\textbf{Network Statistics}} \footnotesize
    \begin{itemize}\setlength{\itemindent}{0pt}
    \item Point statistics
    \begin{itemize}\setlength{\itemindent}{0pt}
    \item Size of nodes and edges
    \item Proportion of nodes in largest connected components
    \item Average degree 
    \item (Tail) Power-law exponent (N/A for non power law networks)
    \item Spectral radius and algebraic connectivity 
    \item Average triangle count
    \item Average clustering coefficient
    \item Degree assortativity coefficient
    \item Max k-core
\end{itemize}
\item  Distributions of network statistics and attributes
\begin{itemize}\setlength{\itemindent}{0pt}
    \item Degree distribution
    \item Pagerank distribution
    \item Singular value distribution (or eigenvalue distribution)
    \item Clustering distribution
    \item Node attribute(s) distribution
    \item Edge attributes(s) distribution
\end{itemize}

\end{itemize}
\end{multicols}
\end{mdframed}
\caption{Network Report Template}
\label{tab:template}
\vspace{-0.2cm}
\end{figure*}

\normalsize

%% file: main.bbl

\begin{thebibliography}{61}


\ifx \showCODEN    \undefined \def \showCODEN     #1{\unskip}     \fi
\ifx \showDOI      \undefined \def \showDOI       #1{#1}\fi
\ifx \showISBNx    \undefined \def \showISBNx     #1{\unskip}     \fi
\ifx \showISBNxiii \undefined \def \showISBNxiii  #1{\unskip}     \fi
\ifx \showISSN     \undefined \def \showISSN      #1{\unskip}     \fi
\ifx \showLCCN     \undefined \def \showLCCN      #1{\unskip}     \fi
\ifx \shownote     \undefined \def \shownote      #1{#1}          \fi
\ifx \showarticletitle \undefined \def \showarticletitle #1{#1}   \fi
\ifx \showURL      \undefined \def \showURL       {\relax}        \fi
\providecommand\bibfield[2]{#2}
\providecommand\bibinfo[2]{#2}
\providecommand\natexlab[1]{#1}
\providecommand\showeprint[2][]{arXiv:#2}

\bibitem[Ahmed et~al\mbox{.}(2013)]%
        {Ahmed2013NetworkSF}
\bibfield{author}{\bibinfo{person}{Nesreen Ahmed}, \bibinfo{person}{Jennifer
  Neville}, {and} \bibinfo{person}{R. Kompella}.}
  \bibinfo{year}{2013}\natexlab{}.
\newblock \showarticletitle{Network Sampling: From Static to Streaming Graphs}.
\newblock \bibinfo{journal}{\emph{ArXiv}}  \bibinfo{volume}{abs/1211.3412}
  (\bibinfo{year}{2013}).
\newblock


\bibitem[Barabsi(2016)]%
        {Barabsi2016NetworkS}
\bibfield{author}{\bibinfo{person}{Albert-Lszl Barabsi}.}
  \bibinfo{year}{2016}\natexlab{}.
\newblock \showarticletitle{Network Science}.
\newblock


\bibitem[Bender(2011)]%
        {Bender2011OnAA}
\bibfield{author}{\bibinfo{person}{Emily~M. Bender}.}
  \bibinfo{year}{2011}\natexlab{}.
\newblock \showarticletitle{On Achieving and Evaluating Language-Independence
  in NLP}.
\newblock \bibinfo{journal}{\emph{Linguistic Issues in Language Technology}}
  \bibinfo{volume}{6} (\bibinfo{year}{2011}).
\newblock


\bibitem[Bender and Friedman(2018)]%
        {Bender2018DataSF}
\bibfield{author}{\bibinfo{person}{Emily~M. Bender} {and}
  \bibinfo{person}{Batya Friedman}.} \bibinfo{year}{2018}\natexlab{}.
\newblock \showarticletitle{Data Statements for Natural Language Processing:
  Toward Mitigating System Bias and Enabling Better Science}.
\newblock \bibinfo{journal}{\emph{Transactions of the Association for
  Computational Linguistics}}  \bibinfo{volume}{6} (\bibinfo{year}{2018}),
  \bibinfo{pages}{587--604}.
\newblock


\bibitem[Beretta et~al\mbox{.}(2018)]%
        {Beretta2018EthicalAS}
\bibfield{author}{\bibinfo{person}{E. Beretta}, \bibinfo{person}{Antonio
  Vetr{\`o}}, \bibinfo{person}{B. Lepri}, {and} \bibinfo{person}{Juan Carlos~De
  Martin}.} \bibinfo{year}{2018}\natexlab{}.
\newblock \showarticletitle{Ethical and Socially-Aware Data Labels}. In
  \bibinfo{booktitle}{\emph{SIMBig}}.
\newblock


\bibitem[Bernard and Killworth(1977)]%
        {Bernard1977INFORMANTAI}
\bibfield{author}{\bibinfo{person}{H. Bernard} {and} \bibinfo{person}{P.
  Killworth}.} \bibinfo{year}{1977}\natexlab{}.
\newblock \showarticletitle{INFORMANT ACCURACY IN SOCIAL NETWORK DATA II}.
\newblock \bibinfo{journal}{\emph{Human Communication Research}}
  \bibinfo{volume}{4} (\bibinfo{year}{1977}), \bibinfo{pages}{3--18}.
\newblock


\bibitem[Blommaert et~al\mbox{.}(2020)]%
        {Blommaert2020TheGG}
\bibfield{author}{\bibinfo{person}{L. Blommaert}, \bibinfo{person}{Roza
  Meuleman}, \bibinfo{person}{Stefan Leenheer}, {and} \bibinfo{person}{Anete
  Butkevica}.} \bibinfo{year}{2020}\natexlab{}.
\newblock \showarticletitle{The gender gap in job authority: Do social network
  resources matter?}
\newblock \bibinfo{journal}{\emph{Acta Sociologica}}  \bibinfo{volume}{63}
  (\bibinfo{year}{2020}), \bibinfo{pages}{381 -- 399}.
\newblock


\bibitem[Blondel et~al\mbox{.}(2008)]%
        {Blondel2008FastUO}
\bibfield{author}{\bibinfo{person}{V. Blondel}, \bibinfo{person}{Jean-Loup
  Guillaume}, \bibinfo{person}{R. Lambiotte}, {and} \bibinfo{person}{E.
  Lefebvre}.} \bibinfo{year}{2008}\natexlab{}.
\newblock \showarticletitle{Fast unfolding of communities in large networks}.
\newblock \bibinfo{journal}{\emph{Journal of Statistical Mechanics: Theory and
  Experiment}}  \bibinfo{volume}{2008} (\bibinfo{year}{2008}),
  \bibinfo{pages}{10008}.
\newblock


\bibitem[Bose and Hamilton(2019)]%
        {Bose2019CompositionalFC}
\bibfield{author}{\bibinfo{person}{A. Bose} {and} \bibinfo{person}{William~L.
  Hamilton}.} \bibinfo{year}{2019}\natexlab{}.
\newblock \showarticletitle{Compositional Fairness Constraints for Graph
  Embeddings}. In \bibinfo{booktitle}{\emph{ICML}}.
\newblock


\bibitem[Buolamwini and Gebru(2018)]%
        {Buolamwini2018GenderSI}
\bibfield{author}{\bibinfo{person}{Joy Buolamwini} {and}
  \bibinfo{person}{Timnit Gebru}.} \bibinfo{year}{2018}\natexlab{}.
\newblock \showarticletitle{Gender Shades: Intersectional Accuracy Disparities
  in Commercial Gender Classification}. In \bibinfo{booktitle}{\emph{FAT}}.
\newblock


\bibitem[Cheng et~al\mbox{.}(2018)]%
        {Cheng2018DataQA}
\bibfield{author}{\bibinfo{person}{Hongju Cheng}, \bibinfo{person}{Da quan
  Feng}, \bibinfo{person}{Xiaobin Shi}, {and} \bibinfo{person}{Chongcheng
  Chen}.} \bibinfo{year}{2018}\natexlab{}.
\newblock \showarticletitle{Data quality analysis and cleaning strategy for
  wireless sensor networks}.
\newblock \bibinfo{journal}{\emph{EURASIP Journal on Wireless Communications
  and Networking}}  \bibinfo{volume}{2018} (\bibinfo{year}{2018}),
  \bibinfo{pages}{1--11}.
\newblock


\bibitem[Eswaran et~al\mbox{.}(2020)]%
        {Eswaran2020HigherOrderLH}
\bibfield{author}{\bibinfo{person}{Dhivya Eswaran}, \bibinfo{person}{Srijan
  Kumar}, {and} \bibinfo{person}{Christos Faloutsos}.}
  \bibinfo{year}{2020}\natexlab{}.
\newblock \showarticletitle{Higher-Order Label Homogeneity and Spreading in
  Graphs}.
\newblock \bibinfo{journal}{\emph{Proceedings of The Web Conference 2020}}
  (\bibinfo{year}{2020}).
\newblock


\bibitem[Fan et~al\mbox{.}(2007)]%
        {Fan2007TheEO}
\bibfield{author}{\bibinfo{person}{Ying Fan}, \bibinfo{person}{M. Li},
  \bibinfo{person}{P. Zhang}, \bibinfo{person}{Jinshan Wu}, {and}
  \bibinfo{person}{Z. Di}.} \bibinfo{year}{2007}\natexlab{}.
\newblock \showarticletitle{The effect of weight on community structure of
  networks}.
\newblock \bibinfo{journal}{\emph{Physica A-statistical Mechanics and Its
  Applications}}  \bibinfo{volume}{378} (\bibinfo{year}{2007}),
  \bibinfo{pages}{583--590}.
\newblock


\bibitem[Fisher(2019)]%
        {Fisher2019MeasuringSB}
\bibfield{author}{\bibinfo{person}{Joseph Fisher}.}
  \bibinfo{year}{2019}\natexlab{}.
\newblock \showarticletitle{Measuring Social Bias in Knowledge Graph
  Embeddings}.
\newblock \bibinfo{journal}{\emph{ArXiv}}  \bibinfo{volume}{abs/1912.02761}
  (\bibinfo{year}{2019}).
\newblock


\bibitem[Freitas et~al\mbox{.}(2021)]%
        {Freitas2021GraphVA}
\bibfield{author}{\bibinfo{person}{Scott Freitas}, \bibinfo{person}{Diyi Yang},
  \bibinfo{person}{Srijan Kumar}, \bibinfo{person}{Hanghang Tong}, {and}
  \bibinfo{person}{Duen~Horng Chau}.} \bibinfo{year}{2021}\natexlab{}.
\newblock \showarticletitle{Graph Vulnerability and Robustness: A Survey}.
\newblock \bibinfo{journal}{\emph{ArXiv}}  \bibinfo{volume}{abs/2105.00419}
  (\bibinfo{year}{2021}).
\newblock


\bibitem[Gebru et~al\mbox{.}(2018)]%
        {Gebru2018DatasheetsFD}
\bibfield{author}{\bibinfo{person}{Timnit Gebru}, \bibinfo{person}{Jamie~H.
  Morgenstern}, \bibinfo{person}{Briana Vecchione},
  \bibinfo{person}{Jennifer~Wortman Vaughan}, \bibinfo{person}{H. Wallach},
  \bibinfo{person}{Hal Daum{\'e}}, {and} \bibinfo{person}{Kate Crawford}.}
  \bibinfo{year}{2018}\natexlab{}.
\newblock \showarticletitle{Datasheets for Datasets}.
\newblock \bibinfo{journal}{\emph{ArXiv}}  \bibinfo{volume}{abs/1803.09010}
  (\bibinfo{year}{2018}).
\newblock


\bibitem[Geiger et~al\mbox{.}(2020)]%
        {Geiger2020GarbageIG}
\bibfield{author}{\bibinfo{person}{R. Geiger}, \bibinfo{person}{Kevin Yu},
  \bibinfo{person}{Yanlai Yang}, \bibinfo{person}{Mindy Dai},
  \bibinfo{person}{Jie Qiu}, \bibinfo{person}{Rebekah Tang}, {and}
  \bibinfo{person}{Jenny Huang}.} \bibinfo{year}{2020}\natexlab{}.
\newblock \showarticletitle{Garbage in, garbage out?: do machine learning
  application papers in social computing report where human-labeled training
  data comes from?}
\newblock \bibinfo{journal}{\emph{Proceedings of the 2020 Conference on
  Fairness, Accountability, and Transparency}} (\bibinfo{year}{2020}).
\newblock


\bibitem[Gjoka et~al\mbox{.}(2010)]%
        {Gjoka2010WalkingIF}
\bibfield{author}{\bibinfo{person}{Minas Gjoka}, \bibinfo{person}{M. Kurant},
  \bibinfo{person}{C. Butts}, {and} \bibinfo{person}{A. Markopoulou}.}
  \bibinfo{year}{2010}\natexlab{}.
\newblock \showarticletitle{Walking in Facebook: A Case Study of Unbiased
  Sampling of OSNs}.
\newblock \bibinfo{journal}{\emph{2010 Proceedings IEEE INFOCOM}}
  (\bibinfo{year}{2010}), \bibinfo{pages}{1--9}.
\newblock


\bibitem[Hameed and Hasso(2020)]%
        {Hameed2020DataPA}
\bibfield{author}{\bibinfo{person}{M. Hameed} {and} \bibinfo{person}{Hasso}.}
  \bibinfo{year}{2020}\natexlab{}.
\newblock \showarticletitle{Data Preparation: A Survey of Commercial Tools}.
\newblock


\bibitem[Hind et~al\mbox{.}(2019)]%
        {Hind2019IncreasingTI}
\bibfield{author}{\bibinfo{person}{M. Hind}, \bibinfo{person}{S. Mehta},
  \bibinfo{person}{A. Mojsilovic}, \bibinfo{person}{R. Nair},
  \bibinfo{person}{K. Ramamurthy}, \bibinfo{person}{Alexandra Olteanu}, {and}
  \bibinfo{person}{K. Varshney}.} \bibinfo{year}{2019}\natexlab{}.
\newblock \showarticletitle{Increasing Trust in AI Services through Supplier's
  Declarations of Conformity}.
\newblock \bibinfo{journal}{\emph{IBM J. Res. Dev.}}  \bibinfo{volume}{63}
  (\bibinfo{year}{2019}), \bibinfo{pages}{6:1--6:13}.
\newblock


\bibitem[Holland et~al\mbox{.}(2018)]%
        {Holland2018TheDN}
\bibfield{author}{\bibinfo{person}{Sarah Holland}, \bibinfo{person}{A. Hosny},
  \bibinfo{person}{Sarah Newman}, \bibinfo{person}{Joshua Joseph}, {and}
  \bibinfo{person}{Kasia Chmielinski}.} \bibinfo{year}{2018}\natexlab{}.
\newblock \showarticletitle{The Dataset Nutrition Label: A Framework To Drive
  Higher Data Quality Standards}.
\newblock \bibinfo{journal}{\emph{ArXiv}}  \bibinfo{volume}{abs/1805.03677}
  (\bibinfo{year}{2018}).
\newblock


\bibitem[Hu et~al\mbox{.}(2020)]%
        {Hu2020OpenGB}
\bibfield{author}{\bibinfo{person}{Weihua Hu}, \bibinfo{person}{Matthias Fey},
  \bibinfo{person}{M. Zitnik}, \bibinfo{person}{Yuxiao Dong},
  \bibinfo{person}{H. Ren}, \bibinfo{person}{Bowen Liu},
  \bibinfo{person}{Michele Catasta}, {and} \bibinfo{person}{J. Leskovec}.}
  \bibinfo{year}{2020}\natexlab{}.
\newblock \showarticletitle{Open Graph Benchmark: Datasets for Machine Learning
  on Graphs}.
\newblock \bibinfo{journal}{\emph{ArXiv}}  \bibinfo{volume}{abs/2005.00687}
  (\bibinfo{year}{2020}).
\newblock


\bibitem[Hutchinson et~al\mbox{.}(2021)]%
        {Hutchinson2021TowardsAF}
\bibfield{author}{\bibinfo{person}{B. Hutchinson}, \bibinfo{person}{A. Smart},
  \bibinfo{person}{A. Hanna}, \bibinfo{person}{Emily~L. Denton},
  \bibinfo{person}{Christina Greer}, \bibinfo{person}{Oddur Kjartansson},
  \bibinfo{person}{P. Barnes}, {and} \bibinfo{person}{Margaret Mitchell}.}
  \bibinfo{year}{2021}\natexlab{}.
\newblock \showarticletitle{Towards Accountability for Machine Learning
  Datasets: Practices from Software Engineering and Infrastructure}.
\newblock \bibinfo{journal}{\emph{Proceedings of the 2021 ACM Conference on
  Fairness, Accountability, and Transparency}} (\bibinfo{year}{2021}).
\newblock


\bibitem[Joyce and Herrmann(2018)]%
        {Joyce2018ARO}
\bibfield{author}{\bibinfo{person}{T. Joyce} {and} \bibinfo{person}{J.
  Herrmann}.} \bibinfo{year}{2018}\natexlab{}.
\newblock \showarticletitle{A Review of No Free Lunch Theorems, and Their
  Implications for Metaheuristic Optimisation}.
\newblock


\bibitem[Jumper et~al\mbox{.}(2021)]%
        {Jumper2021HighlyAP}
\bibfield{author}{\bibinfo{person}{J. Jumper}, \bibinfo{person}{Richard Evans},
  \bibinfo{person}{A. Pritzel}, \bibinfo{person}{Tim Green},
  \bibinfo{person}{Michael Figurnov}, \bibinfo{person}{O. Ronneberger},
  \bibinfo{person}{Kathryn Tunyasuvunakool}, \bibinfo{person}{Russ Bates},
  \bibinfo{person}{Augustin Z{\'i}dek}, \bibinfo{person}{Anna Potapenko},
  \bibinfo{person}{A. Bridgland}, \bibinfo{person}{Clemens Meyer},
  \bibinfo{person}{Simon A~A Kohl}, \bibinfo{person}{Andy Ballard},
  \bibinfo{person}{A. Cowie}, \bibinfo{person}{B. Romera-Paredes},
  \bibinfo{person}{Stanislav Nikolov}, \bibinfo{person}{Rishub Jain},
  \bibinfo{person}{J. Adler}, \bibinfo{person}{T. Back}, \bibinfo{person}{Stig
  Petersen}, \bibinfo{person}{D. Reiman}, \bibinfo{person}{Ellen Clancy},
  \bibinfo{person}{Michal Zielinski}, \bibinfo{person}{Martin Steinegger},
  \bibinfo{person}{Michalina Pacholska}, \bibinfo{person}{Tamas Berghammer},
  \bibinfo{person}{S. Bodenstein}, \bibinfo{person}{D. Silver},
  \bibinfo{person}{Oriol Vinyals}, \bibinfo{person}{A. Senior},
  \bibinfo{person}{K. Kavukcuoglu}, \bibinfo{person}{P. Kohli}, {and}
  \bibinfo{person}{D. Hassabis}.} \bibinfo{year}{2021}\natexlab{}.
\newblock \showarticletitle{Highly accurate protein structure prediction with
  AlphaFold.}
\newblock \bibinfo{journal}{\emph{Nature}} (\bibinfo{year}{2021}).
\newblock


\bibitem[Kaufman et~al\mbox{.}(2011)]%
        {Kaufman2011LeakageID}
\bibfield{author}{\bibinfo{person}{Shachar Kaufman}, \bibinfo{person}{S.
  Rosset}, {and} \bibinfo{person}{Claudia Perlich}.}
  \bibinfo{year}{2011}\natexlab{}.
\newblock \showarticletitle{Leakage in data mining: formulation, detection, and
  avoidance}. In \bibinfo{booktitle}{\emph{KDD}}.
\newblock


\bibitem[Kipf and Welling(2017)]%
        {Kipf2017SemiSupervisedCW}
\bibfield{author}{\bibinfo{person}{Thomas Kipf} {and} \bibinfo{person}{M.
  Welling}.} \bibinfo{year}{2017}\natexlab{}.
\newblock \showarticletitle{Semi-Supervised Classification with Graph
  Convolutional Networks}.
\newblock \bibinfo{journal}{\emph{ArXiv}}  \bibinfo{volume}{abs/1609.02907}
  (\bibinfo{year}{2017}).
\newblock


\bibitem[Koren(2009)]%
        {Koren2009CollaborativeFW}
\bibfield{author}{\bibinfo{person}{Y. Koren}.} \bibinfo{year}{2009}\natexlab{}.
\newblock \showarticletitle{Collaborative filtering with temporal dynamics}. In
  \bibinfo{booktitle}{\emph{KDD}}.
\newblock


\bibitem[Kurant et~al\mbox{.}(2010)]%
        {Kurant2010OnTB}
\bibfield{author}{\bibinfo{person}{M. Kurant}, \bibinfo{person}{A.
  Markopoulou}, {and} \bibinfo{person}{P. Thiran}.}
  \bibinfo{year}{2010}\natexlab{}.
\newblock \showarticletitle{On the bias of BFS (Breadth First Search)}.
\newblock \bibinfo{journal}{\emph{2010 22nd International Teletraffic Congress
  (lTC 22)}} (\bibinfo{year}{2010}), \bibinfo{pages}{1--8}.
\newblock


\bibitem[Landesberger et~al\mbox{.}(2011)]%
        {Landesberger2011VisualAO}
\bibfield{author}{\bibinfo{person}{T.~V. Landesberger}, \bibinfo{person}{Arjan
  Kuijper}, \bibinfo{person}{T. Schreck}, \bibinfo{person}{J. Kohlhammer},
  \bibinfo{person}{J.~V. Wijk}, \bibinfo{person}{Jean-Daniel Fekete}, {and}
  \bibinfo{person}{D. Fellner}.} \bibinfo{year}{2011}\natexlab{}.
\newblock \showarticletitle{Visual Analysis of Large Graphs:
  State‐of‐the‐Art and Future Research Challenges}.
\newblock \bibinfo{journal}{\emph{Computer Graphics Forum}}
  \bibinfo{volume}{30} (\bibinfo{year}{2011}).
\newblock


\bibitem[Leskovec and Faloutsos(2006)]%
        {Leskovec2006SamplingFL}
\bibfield{author}{\bibinfo{person}{J. Leskovec} {and} \bibinfo{person}{C.
  Faloutsos}.} \bibinfo{year}{2006}\natexlab{}.
\newblock \showarticletitle{Sampling from large graphs}. In
  \bibinfo{booktitle}{\emph{KDD '06}}.
\newblock


\bibitem[Leskovec et~al\mbox{.}(2007)]%
        {Leskovec2007GraphED}
\bibfield{author}{\bibinfo{person}{J. Leskovec}, \bibinfo{person}{J.
  Kleinberg}, {and} \bibinfo{person}{C. Faloutsos}.}
  \bibinfo{year}{2007}\natexlab{}.
\newblock \showarticletitle{Graph evolution: Densification and shrinking
  diameters}.
\newblock \bibinfo{journal}{\emph{ACM Trans. Knowl. Discov. Data}}
  \bibinfo{volume}{1} (\bibinfo{year}{2007}), \bibinfo{pages}{2}.
\newblock


\bibitem[Lin et~al\mbox{.}(2014)]%
        {Lin2014MicrosoftCC}
\bibfield{author}{\bibinfo{person}{Tsung-Yi Lin}, \bibinfo{person}{M. Maire},
  \bibinfo{person}{Serge~J. Belongie}, \bibinfo{person}{James Hays},
  \bibinfo{person}{P. Perona}, \bibinfo{person}{D. Ramanan},
  \bibinfo{person}{Piotr Doll{\'a}r}, {and} \bibinfo{person}{C.~L. Zitnick}.}
  \bibinfo{year}{2014}\natexlab{}.
\newblock \showarticletitle{Microsoft COCO: Common Objects in Context}. In
  \bibinfo{booktitle}{\emph{ECCV}}.
\newblock


\bibitem[Lohr(1999)]%
        {Lohr1999SamplingDA}
\bibfield{author}{\bibinfo{person}{S. Lohr}.} \bibinfo{year}{1999}\natexlab{}.
\newblock \showarticletitle{Sampling: Design and Analysis}.
\newblock


\bibitem[Madaio et~al\mbox{.}(2020)]%
        {Madaio2020CoDesigningCT}
\bibfield{author}{\bibinfo{person}{Michael~A. Madaio}, \bibinfo{person}{Luke
  Stark}, \bibinfo{person}{Jennifer~Wortman Vaughan}, {and} \bibinfo{person}{H.
  Wallach}.} \bibinfo{year}{2020}\natexlab{}.
\newblock \showarticletitle{Co-Designing Checklists to Understand
  Organizational Challenges and Opportunities around Fairness in AI}.
\newblock \bibinfo{journal}{\emph{Proceedings of the 2020 CHI Conference on
  Human Factors in Computing Systems}} (\bibinfo{year}{2020}).
\newblock


\bibitem[Marsden(1990)]%
        {Marsden1990NETWORKDA}
\bibfield{author}{\bibinfo{person}{P.~V. Marsden}.}
  \bibinfo{year}{1990}\natexlab{}.
\newblock \showarticletitle{NETWORK DATA AND MEASUREMENT}.
\newblock \bibinfo{journal}{\emph{Review of Sociology}}  \bibinfo{volume}{16}
  (\bibinfo{year}{1990}), \bibinfo{pages}{435--463}.
\newblock


\bibitem[Miceli et~al\mbox{.}(2021)]%
        {Miceli2021DocumentingCV}
\bibfield{author}{\bibinfo{person}{Milagros Miceli}, \bibinfo{person}{Tianling
  Yang}, \bibinfo{person}{Laurens Naudts}, \bibinfo{person}{M. Schuessler},
  \bibinfo{person}{Diana Serbanescu}, {and} \bibinfo{person}{A. Hanna}.}
  \bibinfo{year}{2021}\natexlab{}.
\newblock \showarticletitle{Documenting Computer Vision Datasets: An Invitation
  to Reflexive Data Practices}.
\newblock \bibinfo{journal}{\emph{Proceedings of the 2021 ACM Conference on
  Fairness, Accountability, and Transparency}} (\bibinfo{year}{2021}).
\newblock


\bibitem[Mislove et~al\mbox{.}(2007)]%
        {Mislove2007MeasurementAA}
\bibfield{author}{\bibinfo{person}{A. Mislove}, \bibinfo{person}{M. Marcon},
  \bibinfo{person}{K. Gummadi}, \bibinfo{person}{P. Druschel}, {and}
  \bibinfo{person}{Bobby Bhattacharjee}.} \bibinfo{year}{2007}\natexlab{}.
\newblock \showarticletitle{Measurement and analysis of online social
  networks}. In \bibinfo{booktitle}{\emph{IMC '07}}.
\newblock


\bibitem[Mitchell et~al\mbox{.}(2019)]%
        {Mitchell2019ModelCF}
\bibfield{author}{\bibinfo{person}{Margaret Mitchell}, \bibinfo{person}{Simone
  Wu}, \bibinfo{person}{Andrew Zaldivar}, \bibinfo{person}{P. Barnes},
  \bibinfo{person}{Lucy Vasserman}, \bibinfo{person}{B. Hutchinson},
  \bibinfo{person}{Elena Spitzer}, \bibinfo{person}{Inioluwa~Deborah Raji},
  {and} \bibinfo{person}{Timnit Gebru}.} \bibinfo{year}{2019}\natexlab{}.
\newblock \showarticletitle{Model Cards for Model Reporting}.
\newblock \bibinfo{journal}{\emph{Proceedings of the Conference on Fairness,
  Accountability, and Transparency}} (\bibinfo{year}{2019}).
\newblock


\bibitem[Morris et~al\mbox{.}(2020)]%
        {Morris2020TUDatasetAC}
\bibfield{author}{\bibinfo{person}{Christopher Morris},
  \bibinfo{person}{Nils~M. Kriege}, \bibinfo{person}{Franka Bause},
  \bibinfo{person}{K. Kersting}, \bibinfo{person}{Petra Mutzel}, {and}
  \bibinfo{person}{Marion Neumann}.} \bibinfo{year}{2020}\natexlab{}.
\newblock \showarticletitle{TUDataset: A collection of benchmark datasets for
  learning with graphs}.
\newblock \bibinfo{journal}{\emph{ArXiv}}  \bibinfo{volume}{abs/2007.08663}
  (\bibinfo{year}{2020}).
\newblock


\bibitem[Newman(2003)]%
        {Newman2003TheSA}
\bibfield{author}{\bibinfo{person}{M. Newman}.}
  \bibinfo{year}{2003}\natexlab{}.
\newblock \showarticletitle{The Structure and Function of Complex Networks}.
\newblock \bibinfo{journal}{\emph{SIAM Rev.}}  \bibinfo{volume}{45}
  (\bibinfo{year}{2003}), \bibinfo{pages}{167--256}.
\newblock


\bibitem[Newman(2018)]%
        {newman2018networks}
\bibfield{author}{\bibinfo{person}{Mark Newman}.}
  \bibinfo{year}{2018}\natexlab{}.
\newblock \bibinfo{booktitle}{\emph{Networks}}.
\newblock \bibinfo{publisher}{Oxford university press}.
\newblock


\bibitem[Newman and Watts(1999)]%
        {Newman1999RenormalizationGA}
\bibfield{author}{\bibinfo{person}{M. Newman} {and} \bibinfo{person}{D.
  Watts}.} \bibinfo{year}{1999}\natexlab{}.
\newblock \showarticletitle{Renormalization Group Analysis of the Small-World
  Network Model}.
\newblock \bibinfo{journal}{\emph{Physics Letters A}}  \bibinfo{volume}{263}
  (\bibinfo{year}{1999}), \bibinfo{pages}{341--346}.
\newblock


\bibitem[Ni et~al\mbox{.}(2019)]%
        {Ni2019JustifyingRU}
\bibfield{author}{\bibinfo{person}{Jianmo Ni}, \bibinfo{person}{Jiacheng Li},
  {and} \bibinfo{person}{Julian McAuley}.} \bibinfo{year}{2019}\natexlab{}.
\newblock \showarticletitle{Justifying Recommendations using Distantly-Labeled
  Reviews and Fine-Grained Aspects}. In
  \bibinfo{booktitle}{\emph{EMNLP/IJCNLP}}.
\newblock


\bibitem[Page et~al\mbox{.}(1999)]%
        {Page1999ThePC}
\bibfield{author}{\bibinfo{person}{Lawrence Page}, \bibinfo{person}{S. Brin},
  \bibinfo{person}{R. Motwani}, {and} \bibinfo{person}{T. Winograd}.}
  \bibinfo{year}{1999}\natexlab{}.
\newblock \showarticletitle{The PageRank Citation Ranking : Bringing Order to
  the Web}. In \bibinfo{booktitle}{\emph{WWW 1999}}.
\newblock


\bibitem[Passive~Components(2007)]%
        {miniatureAluminum}
\bibfield{author}{\bibinfo{person}{XICON Passive~Components}.}
  \bibinfo{year}{2007}\natexlab{}.
\newblock \bibinfo{booktitle}{\emph{Miniature Aluminum Electrolytic Capacitors
  XRL Series}}.
\newblock
\urldef\tempurl%
\url{https://www.mouser.com/catalog/specsheets/XC-600178.pdf}
\showURL{%
\tempurl}


\bibitem[Rubinov and Sporns(2010)]%
        {Rubinov2010ComplexNM}
\bibfield{author}{\bibinfo{person}{M. Rubinov} {and} \bibinfo{person}{O.
  Sporns}.} \bibinfo{year}{2010}\natexlab{}.
\newblock \showarticletitle{Complex network measures of brain connectivity:
  Uses and interpretations}.
\newblock \bibinfo{journal}{\emph{NeuroImage}}  \bibinfo{volume}{52}
  (\bibinfo{year}{2010}), \bibinfo{pages}{1059--1069}.
\newblock


\bibitem[Safavi and Koutra(2020)]%
        {Safavi2020CoDExAC}
\bibfield{author}{\bibinfo{person}{Tara Safavi} {and} \bibinfo{person}{Danai
  Koutra}.} \bibinfo{year}{2020}\natexlab{}.
\newblock \showarticletitle{CoDEx: A Comprehensive Knowledge Graph Completion
  Benchmark}. In \bibinfo{booktitle}{\emph{EMNLP}}.
\newblock


\bibitem[Shin et~al\mbox{.}(2016)]%
        {Shin2016CoreScopeGM}
\bibfield{author}{\bibinfo{person}{Kijung Shin}, \bibinfo{person}{Tina
  Eliassi-Rad}, {and} \bibinfo{person}{C. Faloutsos}.}
  \bibinfo{year}{2016}\natexlab{}.
\newblock \showarticletitle{CoreScope: Graph Mining Using k-Core Analysis —
  Patterns, Anomalies and Algorithms}.
\newblock \bibinfo{journal}{\emph{2016 IEEE 16th International Conference on
  Data Mining (ICDM)}} (\bibinfo{year}{2016}), \bibinfo{pages}{469--478}.
\newblock


\bibitem[Stoica et~al\mbox{.}(2018)]%
        {Stoica2018AlgorithmicGC}
\bibfield{author}{\bibinfo{person}{Ana-Andreea Stoica},
  \bibinfo{person}{Christopher~J. Riederer}, {and} \bibinfo{person}{A.
  Chaintreau}.} \bibinfo{year}{2018}\natexlab{}.
\newblock \showarticletitle{Algorithmic Glass Ceiling in Social Networks: The
  effects of social recommendations on network diversity}.
\newblock \bibinfo{journal}{\emph{Proceedings of the 2018 World Wide Web
  Conference}} (\bibinfo{year}{2018}).
\newblock


\bibitem[Stumpf et~al\mbox{.}(2005)]%
        {Stumpf2005SubnetsOS}
\bibfield{author}{\bibinfo{person}{M. Stumpf}, \bibinfo{person}{C. Wiuf}, {and}
  \bibinfo{person}{R. May}.} \bibinfo{year}{2005}\natexlab{}.
\newblock \showarticletitle{Subnets of scale-free networks are not scale-free:
  sampling properties of networks.}
\newblock \bibinfo{journal}{\emph{Proceedings of the National Academy of
  Sciences of the United States of America}}  \bibinfo{volume}{102 12}
  (\bibinfo{year}{2005}), \bibinfo{pages}{4221--4}.
\newblock


\bibitem[Takac(2012)]%
        {Takac2012DATAAI}
\bibfield{author}{\bibinfo{person}{Lubos Takac}.}
  \bibinfo{year}{2012}\natexlab{}.
\newblock \showarticletitle{DATA ANALYSIS IN PUBLIC SOCIAL NETWORKS}.
\newblock


\bibitem[Toutanova et~al\mbox{.}(2015)]%
        {Toutanova2015RepresentingTF}
\bibfield{author}{\bibinfo{person}{Kristina Toutanova}, \bibinfo{person}{Danqi
  Chen}, \bibinfo{person}{P. Pantel}, \bibinfo{person}{Hoifung Poon},
  \bibinfo{person}{Pallavi Choudhury}, {and} \bibinfo{person}{M. Gamon}.}
  \bibinfo{year}{2015}\natexlab{}.
\newblock \showarticletitle{Representing Text for Joint Embedding of Text and
  Knowledge Bases}. In \bibinfo{booktitle}{\emph{EMNLP}}.
\newblock


\bibitem[Traag et~al\mbox{.}(2015)]%
        {Traag2015DetectingCU}
\bibfield{author}{\bibinfo{person}{V. Traag}, \bibinfo{person}{R. Aldecoa},
  {and} \bibinfo{person}{J. Delvenne}.} \bibinfo{year}{2015}\natexlab{}.
\newblock \showarticletitle{Detecting communities using asymptotical Surprise}.
\newblock \bibinfo{journal}{\emph{Physical review. E, Statistical, nonlinear,
  and soft matter physics}}  \bibinfo{volume}{92 2} (\bibinfo{year}{2015}),
  \bibinfo{pages}{022816}.
\newblock


\bibitem[Tsourakakis(2010)]%
        {Tsourakakis2010CountingTI}
\bibfield{author}{\bibinfo{person}{Charalampos~E. Tsourakakis}.}
  \bibinfo{year}{2010}\natexlab{}.
\newblock \showarticletitle{Counting triangles in real-world networks using
  projections}.
\newblock \bibinfo{journal}{\emph{Knowledge and Information Systems}}
  \bibinfo{volume}{26} (\bibinfo{year}{2010}), \bibinfo{pages}{501--520}.
\newblock


\bibitem[Wasserman(2004)]%
        {Wasserman2004AllOS}
\bibfield{author}{\bibinfo{person}{L. Wasserman}.}
  \bibinfo{year}{2004}\natexlab{}.
\newblock \showarticletitle{All of Statistics: A Concise Course in Statistical
  Inference}.
\newblock


\bibitem[Watts and Strogatz(1998)]%
        {Watts1998CollectiveDO}
\bibfield{author}{\bibinfo{person}{D. Watts} {and} \bibinfo{person}{S.
  Strogatz}.} \bibinfo{year}{1998}\natexlab{}.
\newblock \showarticletitle{Collective dynamics of ‘small-world’ networks}.
\newblock \bibinfo{journal}{\emph{Nature}}  \bibinfo{volume}{393}
  (\bibinfo{year}{1998}), \bibinfo{pages}{440--442}.
\newblock


\bibitem[Yang et~al\mbox{.}(2018)]%
        {Yang2018ANL}
\bibfield{author}{\bibinfo{person}{Ke Yang}, \bibinfo{person}{Julia
  Stoyanovich}, \bibinfo{person}{Abolfazl Asudeh}, \bibinfo{person}{Bill Howe},
  \bibinfo{person}{H.~V. Jagadish}, {and} \bibinfo{person}{G. Miklau}.}
  \bibinfo{year}{2018}\natexlab{}.
\newblock \showarticletitle{A Nutritional Label for Rankings}.
\newblock \bibinfo{journal}{\emph{Proceedings of the 2018 International
  Conference on Management of Data}} (\bibinfo{year}{2018}).
\newblock


\bibitem[Yao and Huang(2017)]%
        {Yao2017BeyondPF}
\bibfield{author}{\bibinfo{person}{Sirui Yao} {and} \bibinfo{person}{Bert
  Huang}.} \bibinfo{year}{2017}\natexlab{}.
\newblock \showarticletitle{Beyond Parity: Fairness Objectives for
  Collaborative Filtering}. In \bibinfo{booktitle}{\emph{NIPS}}.
\newblock


\bibitem[Ying et~al\mbox{.}(2018)]%
        {Ying2018GraphCN}
\bibfield{author}{\bibinfo{person}{Rex Ying}, \bibinfo{person}{Ruining He},
  \bibinfo{person}{K. Chen}, \bibinfo{person}{Pong Eksombatchai},
  \bibinfo{person}{William~L. Hamilton}, {and} \bibinfo{person}{J. Leskovec}.}
  \bibinfo{year}{2018}\natexlab{}.
\newblock \showarticletitle{Graph Convolutional Neural Networks for Web-Scale
  Recommender Systems}.
\newblock \bibinfo{journal}{\emph{Proceedings of the 24th ACM SIGKDD
  International Conference on Knowledge Discovery \& Data Mining}}
  (\bibinfo{year}{2018}).
\newblock


\bibitem[You et~al\mbox{.}(2018)]%
        {You2018GraphCP}
\bibfield{author}{\bibinfo{person}{Jiaxuan You}, \bibinfo{person}{B. Liu},
  \bibinfo{person}{Rex Ying}, \bibinfo{person}{V. Pande}, {and}
  \bibinfo{person}{J. Leskovec}.} \bibinfo{year}{2018}\natexlab{}.
\newblock \showarticletitle{Graph Convolutional Policy Network for
  Goal-Directed Molecular Graph Generation}. In
  \bibinfo{booktitle}{\emph{NeurIPS}}.
\newblock


\end{thebibliography}
